\documentclass[12pt,preprint]{aastex}

\usepackage{rotating}

\shorttitle{{\it AstroSat/LAXPC} View of GX 17+2}
\shortauthors{Agrawal et al.}

\begin{document}

\title{ 
{\it AstroSat/LAXPC} view of GX 17+2: Spectral Evolution along the Z-track }


\author{V. K. Agrawal\altaffilmark{1}}

\email{vivekag@ursc.gov.in}

\author{Anuj Nandi\altaffilmark{1}}
\email{anuj@ursc.gov.in}
\and
\author{M. C. Ramadevi \altaffilmark{1}}
\email{ramadevi@ursc.gov.in}


\altaffiltext{1}{Space Astronomy Group, ISITE Campus, U. R. Rao Satellite Center, Bangalore, 560037, India }


\begin{abstract}
In this paper, we present the first results obtained using $\sim$ 50
ks observations of the bright low-mass X-ray binary (LMXB) GX 17+2 with
Large Area X-ray Proportional Counter (LAXPC) onboard {\it AstroSat}. The
source traced out a complete Z-track in the hardness intensity diagram
(HID). The spectra at different sections of the Z-diagram are well
described by either a combination of a thermal Comptonization component, a
power-law and a relativistic iron line or a model consisting of  a thermal
disk component, a single temperature blackbody, a power-law and a
relativistic iron line. Fitting the spectra with both phenomenological
models suggests that the power-law component is strong in the horizontal
branch (HB), becomes weaker as the source moves down the normal branch
(NB) and then again becomes stronger as the sources moves up the flaring
branch (FB). However, we find that the strength of the power-law component
is model dependent, although the trend in the variation of the
power-law strength along the Z-track is similar. A simple model
composed by a Comptonized emission and power-law component, convolved
with the ionized reflection,  also describes the spectra very well.

 A normal branch oscillation
(NBO) with a centroid frequency 7.42$\pm$0.23 Hz, quality factor (Q)
$\sim$ 4.88, rms 1.41$\pm$0.29\% and significance 5.1$\sigma$ is detected
at the middle of the NB. 
The parameters of the Comptonized emission show a systematic evolution
along the Z-diagram. The optical depth of the corona increases as the
source moves up along the FB, suggesting possible trigger of an outflow
or dumping of the disc material in to the corona by radiation pressure.
\end{abstract}





\section{Introduction}
Z-sources, a subclass of neutron star low-mass X-ray binaries (NS-LMXBs),
are among the most luminous and persistent X-ray sources in the sky.
They have X-ray luminosities typically in the range of $0.5-1.0$ times the
Eddington limit ($L_{Edd}$) for 1.4 $M_\odot$ neutron star. They are named
so as they trace out a `Z' pattern \citep{Hasvan89} in the color-color
diagram (CCD) and the hardness intensity diagram (HID). From the top
left, three branches of the Z-track are: horizontal branch (HB), normal
branch (NB) and flaring branch (FB). Another subclass of NS-LMXBs is
atoll sources. They trace out a `C-type' pattern in the CCD and the
HID. The transient Z-source XTE J1701-462 \citep{Remillard06} showed both
atoll and Z-source like behaviour at different flux levels, suggesting
that both are similar system accreting at different rates \citep{Homan07,Lin09}. 

Though spectral properties of these sources have been investigated
in detail, results of spectral modelling are not conclusive and model
degeneracy exists. Between two widely accepted scenario, one evokes a
model which is sum of a multicolour blackbody emission from a standard
disc and a component resulting from inverse Compton scattering of the
soft seed photons by hot plasma in the boundary-layer/central-corona
\citep{Mit84,Barret01, disalvo02,agrawal03,agrawal09}. In another
scenario, the emitted spectrum is modeled as sum of a single
temperature blackbody component from  the boundary-layer/NS-surface
and the hard Comptonized component from the hot inner accretion flow
\citep{disalvo00,disalvo01,Sleator16}. The choice of the soft component
between the multicolour disc emission (first scenario) and the blackbody emission
from the neutron star surface (second scenario) makes these two approaches different. Though, the
Comptonized component in both approaches is described by same model, it
may be associated with different parts of the accretion flow. Recent
work by \cite{Lin12} has shown that sum of a multi-colour disc emission
and a blackbody emission also provides acceptable description of the
X-ray spectra of NS-LMXBs. However, this model is more appropriate for the spectra without hard tails or high energy coverage. A non-thermal power-law component has been
reported in the energy spectra of five persistent Z-sources (Cyg X-2,
\citealt{disalvo02}; GX 17+2, \citealt{disalvo00,Farinelli05}; Sco
X-1, \citealt{revni14,disalvo06,dai07, damico1}; GX 5-1,
\citealt{asai94}; GX 349+2, \citealt{disalvo01}; and the peculiar source Cir X-1, \citealt{Iaria01}) and one transient
Z-source XTE J1701-462 \citep{Ding11}. A hint of a power-law component is also present in the BeppoSAX spectrum of the Z-source GX 340+0 \citep{Iaria06}.

Investigations of the fast time variability revealed rich varieties
of Quasi-periodic Oscillations (QPOs) and noise components in
the power density spectra (PDS) of the Z-sources \citep{van95}. These
features are tightly correlated with the positions on the Z-diagram
\citep{jonker00,jonker02,Homan02}. QPOs in the frequency range $15-80$ Hz
are seen in the HB and the upper part of NB. They are called horizontal
branch oscillations (HBOs). QPOs in the frequency range $5-7$ Hz are also seen
in the middle part of NB to the lower part of NB and are called normal branch
oscillations (NBOs). The frequency of NBO suddenly increases to $15-30$
Hz as the source crosses the NB/FB vertex. It is believed that at the
NB/FB vertex NBO is transformed into flaring branch oscillations (FBOs).
A pair of kHz QPOs in the frequency range $200-1200$ Hz have been seen
in the six persistent Z-sources (Sco X-1, \citealt{van96}; Cyg X-2,
\citealt{wij98a}; GX 5-1, \citealt{wij98b}; GX 340+0, \citealt{jonker98};
GX 17+2, \citealt{wij97}; GX 349+2, \citealt{zhang98}) and the
transient Z-source \citep{Sanna10}. The noise components observed in the
Z-sources are: Very low frequency noise (VLFN), low frequency noise
(LFN) and high frequency noise (HFN). LFN and HFN are band limited
noise and are modeled with a cutoff power-law or a Lorentzian component and VLFN
is modeled with a power-law. A smooth evolution of the power spectral
features along the CCD suggests that the energy spectra should also
vary smoothly along the Z-track. Studies have been carried out to
understand the evolution of the X-ray spectral parameters along the
Z-track \citep{disalvo00,disalvo02,done02,agrawal03,agrawal09,Lin12}. There
are also attempts to investigate the correlation between the spectral
parameters and the parameters of the power spectral features \citep{Titar07,Lin12}.

GX 17+2 is a bright and persistent NS-LMXB classified as Z-source
\citep{Hasvan89}. The source has exhibited all types of QPOs (HBO,
NBO, FBO and pair of kHz QPOs) so far identified in the Z-sources. A
detailed temporal evolution along the Z-track has been investigated
using the {\it Rossi X-ray Timing Explorer (RXTE)} data \citep{Homan02}. A
detailed X-ray spectral studies have been carried out previously using the
{\it BeppoSAX} \citep{disalvo00} and {\it RXTE} data \citep{Lin12}. \cite{Lin12}
fitted the X-ray spectra with a model consisting of a single temperature
blackbody ({\it bbodyrad} in XSPEC), a multi-colour disc component ({\it
diskbb} in XSPEC), a weak Comptonized component described by cutoff
powerlaw and a Gaussian line with energy fixed at 6.6 keV. They noted
that frequency of the upper kHz QPO was correlated with the inner disc
radius. {\it BeppoSAX} spectra of the source were fitted with a blackbody and
Comptonized component \citep{disalvo00}. A power-law tail with photon
index $\sim$ 2.7 was also found to be present in the HB spectra of the
source, perhaps correlated with the radio (jet) emission (see, \citealt{mig07}). Further, a signature of reflection feature was also seen during
the {\it NuSTAR} observations of this source \citep{ludham17}. Modelling of the
{\it NuSTAR} spectrum with the reflection model suggested a low inclination
($25-38$$^\circ$) of the system. An inclination angle of $\sim$ 30$^\circ$
for this neutron star system was derived by fitting the relativistic
iron line observed with the {\it Suzaku} data \citep{cackett10}. Analysis of
the {\it NuSTAR} observation \citep{ludham17} also suggested that the accretion
disc is truncated close to the inner most stable circular orbit (ISCO).

In this paper, we report the results obtained by the first {\it AstroSat/LAXPC} 
observations of the source GX 17+2. We present the spectral and temporal 
evolution of the source along the Z-track utilizing broad band ($3-80$ keV) 
data from the LAXPC instrument. The source traced a Z-track in the HID during 
the observations and showed the presence of a QPO in the normal branch. We report 
the detection of a hard power-law tail in both HB and FB branch of the Z-track. The
paper is organized as follows. The observations and data reduction are
described in $\S$2. $\S$3 deals with the methods of the data analysis
and modelling of the temporal and spectral data. The results of spectral
and timing analysis are presented in $\S$4. In section $\S$5, we interpret
our results and conclude.
\section{Observations and Data Reduction} 

The source GX 17+2 was observed from May 11, 2016 to May 14, 2016 with
Large Area X-ray Proportional Counter (LAXPC) onboard {\it AstroSat}.
The data was collected during Guaranteed Time (GT) observation. The source
was observed for a total exposure time of 50 ks. LAXPC instruments provide
high time resolution (10 micro second) and  moderate energy resolution
data in the $3-80$ keV band. Three co-aligned identical proportional
counters (LAXPC10, LAXPC20 and LAXPC30) provide a combined effective area
of $\sim$ 6000 cm$^2$ at 15 keV \citep{yadav16, agr17, antia17}. The
detection efficiency $>$ 50 per cent above 30 keV is achieved by
the LAXPC detection volume, which is a mixture of xenon and methane at
2 atmospheric pressure. For the data reduction, we use the software and
follow the procedures provided in the webpage http://astrosat-ssc.iucaa.in/.
We use recent calibration files and background models provided by
the instrument team (see \cite{antia17} for more details).
\section{Data Analysis}

\subsection{Lightcurve and Z-track}
The background subtracted binned lightcurve in the energy range
$3.0-60.0$ keV is created using the LAXPC10 event mode data. The
average background count rate varies between $170-190$ counts/s in the
energy band of $3.0-60.0$ keV. We use a bin size of 256 sec for the
lightcurve generation. An investigation of the lightcurve reveals that
initially the source intensity is roughly constant ($1900-2000$ cts/s)
and then the source intensity gradually decreases from $\sim$ 1900 cts/s
to $\sim$ 1500 cts/s. This decaying phase is interrupted by a sudden
increase (flare) in the source count rates (see Fig. 1).
\begin{figure}
\centering
\includegraphics[width=0.35\textwidth,angle=-90]{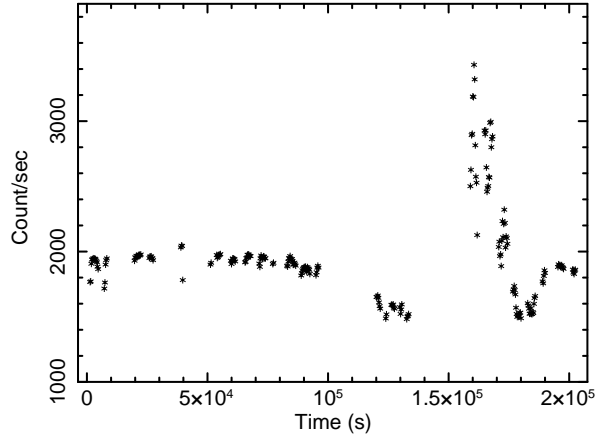}
\caption{Background subtracted lightcurve of GX 17+2 in the energy band $3.0-60.0$ keV. 
The lightcurve is created using 256 sec binsize. Only LAXPC10 has been considered for 
creating the lightcurve. The errors on the data points are smaller than the marker size.}
\end{figure}

Hardness Intensity Diagram (HID) is created using the  background subtracted 
lightcurves in the $3.0-20.0$ keV, $7.3-10.5$ keV and $10.5-20.0$ keV energy 
bands. Here, the hard colour is defined as the ratio
of count rates in the energy bands $10.5-20.0$ keV and $7.3-10.5$ keV. The
energy bands used to  define the hard colour and intensity are similar
to that used by \cite{Homan02} for a comparison. To create the HID, we
plot the hard-colour against the source intensity in the energy band
$3.0-20.0$ keV. The HID is shown in Figure 2. All the three branches of
the Z-track can be identified clearly in the HID. To study the spectral
and temporal evolution along the Z-track, we divide the Z-track into
9 segments. The horizontal branch (HB) is divided into two parts `HB1' and
`HB2'. The normal branch (NB) is divided into three parts `NB1', `NB2' and
`NB3'. The flaring branch (FB) is divided into total 4 divisions, `FB1',
`FB2', `FB3' and `FB4'. These sections are marked in the HID. The top
left part of the HB (HB1) is almost horizontal. However, the HB2 section is
almost vertical and it continues to bend left into the NB. For each of
these sections, we create the X-ray spectra and PDS (Power Density Spectra). These spectra and
PDS are  used for further spectral and temporal analysis.

\begin{figure}
\centering
\includegraphics[width=0.35\textwidth,angle=-90]{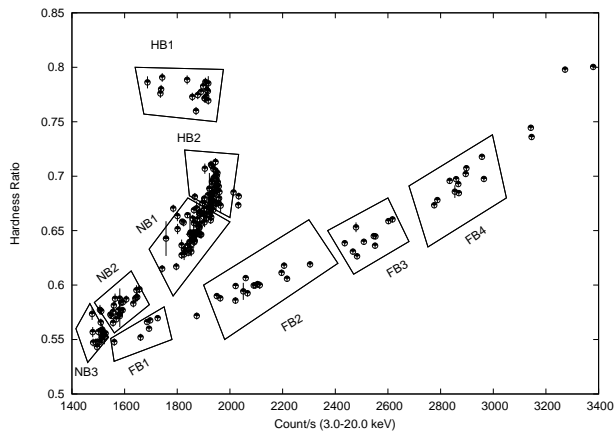}
\caption{Hardness Intensity Diagram (HID) of GX 17+2 created using LAXPC10 data. Each point 
on the HID corresponds to 256 sec binsize. The marked regions show the sections used 
to create the energy spectra and power density spectra. See text for details.}
\end{figure}

\subsection{Spectral Analysis} 

First, we create the source and background spectra for the different
sections of the HID.  The background subtracted spectra are fitted
using the spectral fitting software XSPEC version 12.9.1. Since LAXPC10
is well calibrated and have less background issues, we use data from
this detector only to perform the  spectral fitting. Unless mentioned
explicitly, all the errors are computed using $\Delta \chi^2=1.0$ (68\%
confidence level).  We included 1\% systematics in overall fitting to
account for the uncertainty in the spectral response \citep{agrawal18}.

We attempt multi-component phenomenological spectral models to fit
the X-ray spectra of the source in HB1. First, we fit the data with the
combination of a thermal Comptonization and a broad Gaussian line.
To represent the Comptonized emission, we use {\it nthComp} model of
XSPEC \citep{Zdz96}.  In our model, we assume the source of the seed
photon as blackbody emission.
The fitting gives the reduced $\chi^2$ ($\chi^2/dof$) = 109/81. We find
that there is an excess of emission above 30 keV  on the top of the {\it nthComp}
model. To account for this excess emission, we include a power-law
component which improved the fit significantly ($\chi^2/dof$=66/79)
with a chance improvement probability of 2.5e-9. The parameters of the Gaussian component and reduced $\chi^2$ obtained by fitting the data with {\it wabs(nthcomp+Gauss+power-law)} model are given in Table 2.
\begin{figure*}
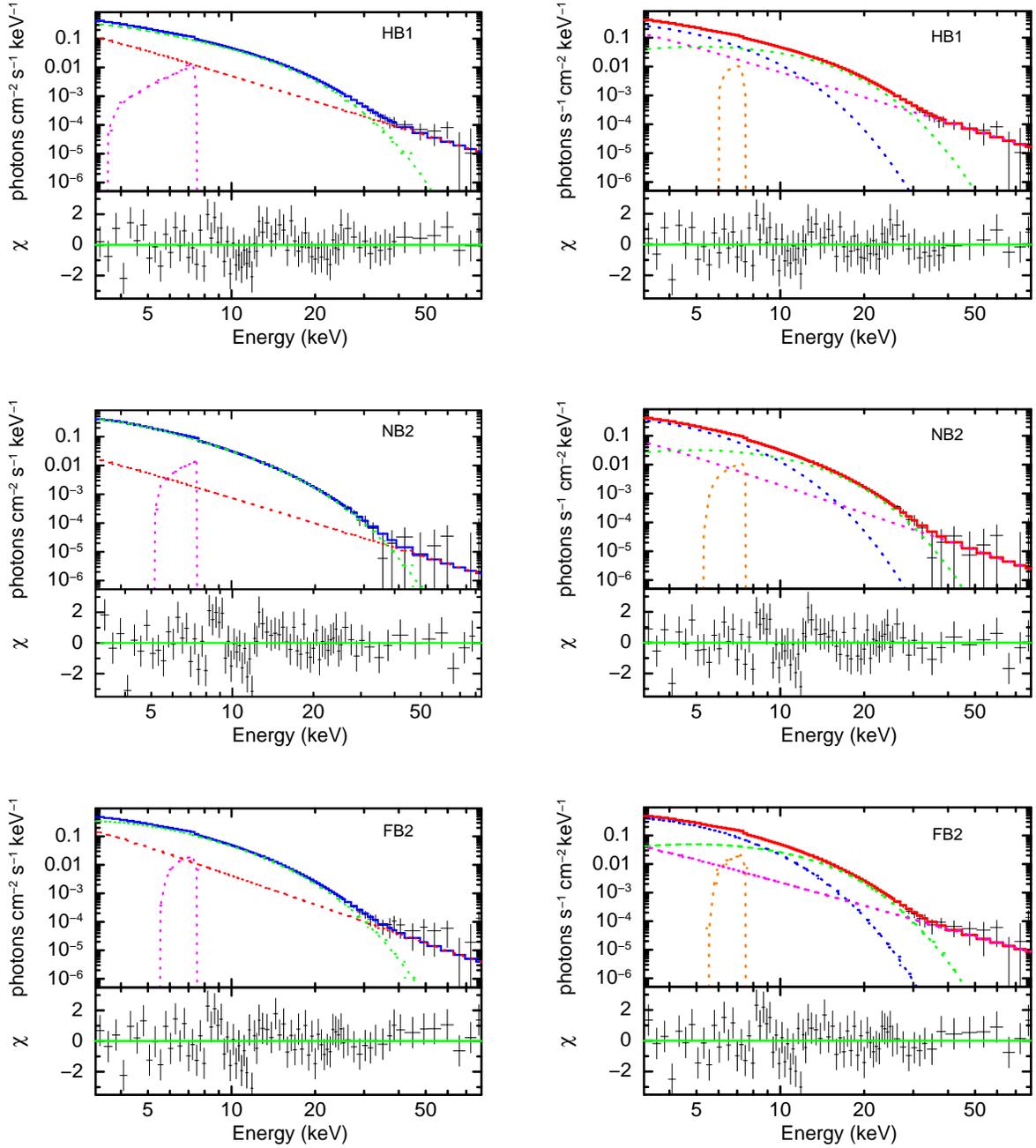

\centering
\begin{tabular}{@{}cc@{}}
\includegraphics[width=0.34\textwidth,angle=-90]{HB1-figure3a-spec-reb.eps}&
\includegraphics[width=0.34\textwidth, angle=-90]{HB1-figure3a-spec-reb-mcd.eps} \\
\includegraphics[width=0.34\textwidth,angle=-90]{NB2-figure3b-spec-reb.eps}&
\includegraphics[width=0.34\textwidth,angle=-90]{NB2-figure3b-spec-reb-mcd.eps}\\
\includegraphics[width=0.34\textwidth,angle=-90]{FB2-figure3c-spec-reb.eps}&
\includegraphics[width=0.34\textwidth,angle=-90]{FB2-figure3c-spec-reb-mcd.eps}
\\
\end{tabular}
\caption{The unfolded spectra {\bf{Model 1:} {\it nthComp+diskline+power-law}} on the left panel and {\bf Model 2:} {\it diskbb+bbodyrad+diskline+power-law} on the right panel of GX 17+2 at three different parts of the Z-track. The top panel shows the spectra at HB1. A clear signature of a hard tail extending above 30 keV is visible in this panel. The bottom panel displays the spectra at FB2 showing the signature of hard tail. The middle panel, which shows the spectra at NB2, suggests that hard tail is weak in the NB.  }
\end{figure*}
\begin{figure*}
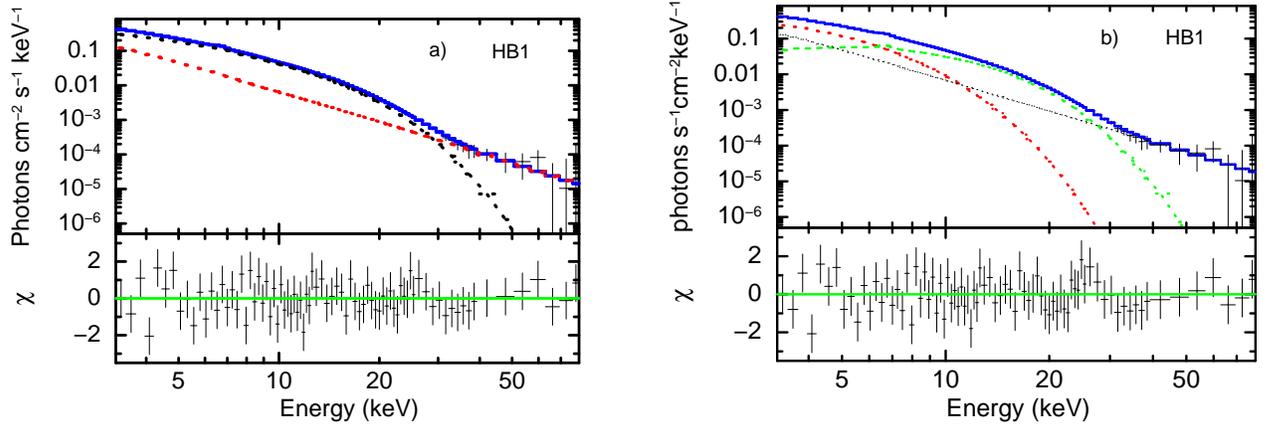

\centering
\begin{tabular}{@{}cc@{}}
\includegraphics[width=0.36\textwidth,angle=-90]{HB1-rfxconv.eps}&
\includegraphics[width=0.36\textwidth, angle=-90]{HB1-bbrefl-relconv.eps} \\
\end{tabular}
\caption{ a) The unfolded spectrum for HB1 section and the used model is {\bf Model 3} : {\it rfxconv*(nthComp+power-law)}. b) The unfloded spectrum for the HB1 and the used model is {\bf Model 4} : {\it diskbb+power-law+relconv*bbrefl}. The residuals in unit of sigma are shown in the bottom panels of a) and b).}
\end{figure*}

\begin{figure*}
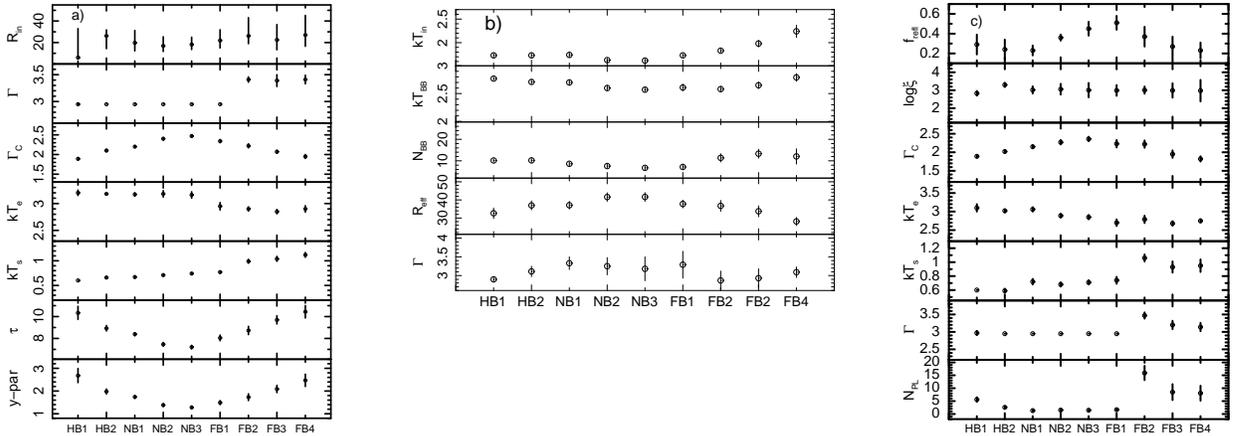

\begin{tabular}{@{}ccc@{}}
\includegraphics[width=0.35\textwidth,angle=-90]{parameter-evolution.eps}
&
\includegraphics[width=0.25\textwidth,angle=-90]{mcd-plot-evolution.eps}&
\includegraphics[width=0.35\textwidth,angle=-90]{rfxconv-plot.eps} \\
\end{tabular}
\caption{Evolution of the best fit spectral parameters along the Z-track of GX 17+2  {\bf a)} for {\bf Model 1}, b) for {\bf Model 2}, and {\bf c) for Model 3}. See text, Table 1, Table 3 and Table 4 for details.}
 \end{figure*}
\begin{figure*}
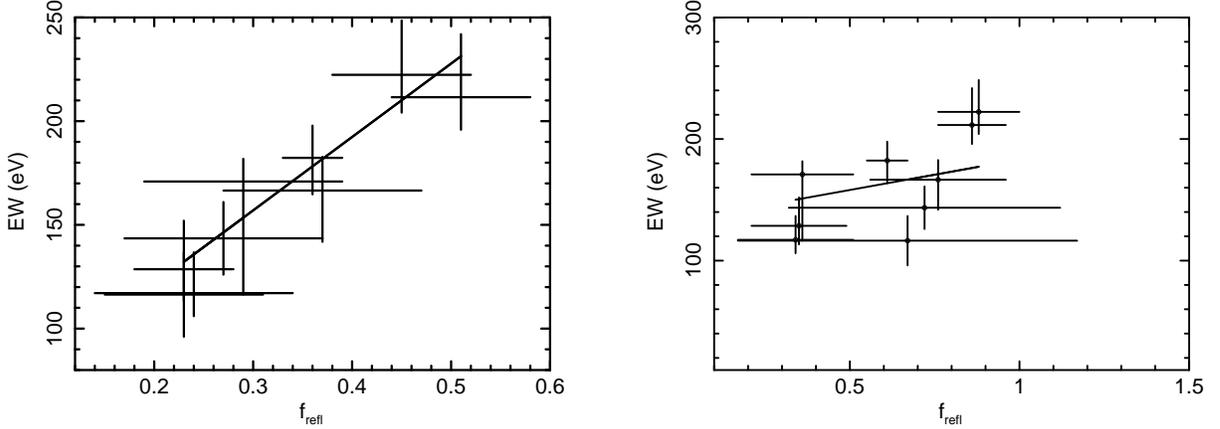

\begin{tabular}{@{}cc@{}}
\includegraphics[width=0.35\textwidth,angle=-90]{correlation-eq-refl.eps}&
\includegraphics[width=0.35\textwidth,angle=-90]{correlation-eq-bbrefl.eps}
\end{tabular}
\caption{ Left: Correlation between the equivalent width of the iron line ( {\bf Model 1})  and the reflection fraction (from {\bf Model 3}). Right: Correlation between the equivalent width of the iron line ({\bf Model 1})  and the reflection fraction (from {\bf Model 4})}
\end{figure*}

We also model the broad Gaussian feature with {\it diskline} model
\citep{Fab89} which takes into account the relativistic effect for
a Schwarzschild space-time geometry. The {\it diskline} model has 6
free parameters: inner disc radius, outer disc radius, line energy,
inclination, emissivity index and normalization. We kept the inner
disc radius, normalization and line energy free. The inclination
angle derived from the {\it NuSTAR} data of this source varies between
$\sim$ $25-40$$^\circ$ \citep{ludham17} and hence we allow to vary the
inclination angle between these values. We find that inclination angle
peges at 40$^\circ$. Hence, we fix the inclination at this value. We
also fix the emissivity index at -3 and outer disc radius at 1000 
gravitational radii. Fitting the iron line feature with {\it diskline}
model, we get a similar reduced $\chi^2$($\chi^2/dof$=67/79) with respect to the fit with a Gaussian profile and hence
we adopt {\it wabs(nthComp+diskline+power-law)} as the best description of
the GX 17+2 spectrum at the section HB1. We use this model to fit the
spectra from other sections of Z-track and find it adequate (see Table
1). We refer to this model as {\bf Model 1}.

We also tried a multi-component model, consisting of emission
from a multi-colour disk, a blackbody emission component from
boundary-layer/NS-surface and a relativistic iron line to fit the spectra
at different sections of the Z-track. A significant power-law component
is also required in order to fit the spectra at different parts of the
Z-track (see Table 3). To model the emission from disc, we use {\it diskbb}
model of XSPEC and to represent the blackbody we use {\it bbodyrad}
model in XSPEC. We find that {\it diskbb+bbodyrad+power-law+diskline
} model also successfully describes the spectra of this source. We refer
to this model as {\bf Model 2}. 
 To understand the nature of iron line we also tried a convolution
 model {\it rfxconv} which takes into account the reflection from
 an ionized disk  \citep{Koleh11}. We refer to the model {\it
 wabs*rfxconv*(nthcomp+power-law)} as {\bf Model 3}. We allow
 the inclination angle to vary between 25$^\circ$ to 40$^\circ$
 (see \cite{ludham17}). The inclination angle peges to the lower
 limit 25$^\circ$ and hence we fix the inclination at 25$^\circ$
 while fitting the spectra with this model. We find that {\bf Model 3}
 provides comparable and even better fit (for some section of the Z-track)
 compared to {\bf Model 1} and {\bf Model 2} (see Table 4). In order to
 compare our results with previous work \citep{ludham17}, we also aply
 BBREFL \citep{Ballantyne04} convolved with RELCONV \citep{Dauser10}
 which describe the reflection of blackbody component from the
 accretion disc. We find that {\it diskbb+relconv*bbrefl+power-law}
 model successfully describes the spectra at different parts of Z-track
 (see Table 5). We refer to this model as {\bf Model 4}. While fitting the
 spectra with this model we allow the inclination to vary between $25^\circ -
 40^\circ$. The inclination peges to the lower limit. We also note that
 if we keep the inclination angle and the inner disk radius as free parameters,
 the inner disk radius is not constrained. Hence, we fix the inclination
 at the lower limit 25$^\circ$.  The neutral hydrogen column density is
 fixed to 2 $\times 10^{22} cm^{-2}$ \citep{disalvo00}  in all the
four models described above.

\subsection{Timing Analysis}
First, we generate the lightcurves with a binsize of 0.4 milli seconds
in the $3.0-60.0$ keV band. Lightcurves are divided into intervals of 32768
bins. For each interval, we create the PDS. The PDS corresponding to the
same section of the HID are averaged and then rebinned in the frequency
space. The binned PDS are normalized to the fractional rms spectra
(in units of $(rms/mean)^2 Hz^{-1}$). Dead time corrected Poisson
noise level (see \citealt{Zhang95}; \citealt{agrawal18}) is subtracted
from all the PDS. Since power is not significant above 100 Hz, we use only
$0.07-100$ Hz frequency range for the fitting. The PDS is fitted with
the combinations of a power-law and/or Lorentzian functions. We find
two types of noise features in the PDS: a very low frequency noise (VLFN)
modelled using power-law function ($A \nu^{-\alpha}$) with index $\alpha$
and normalization $A$ and band limited noise (BLN) component modelled using
Lorentzian function defined as,
\begin{equation}
F_\nu = LN/(1+(2(\nu - \nu_{L})/FWHM)^2),
\end{equation}

where $LN$ is the normalization, $\nu_{L}$ is the centroid frequency
and $FWHM$ is full-width-half-maximum of the Lorentzian. In this
representation, $\pi *LN * FWHM/2$ gives the integral of Lorentzian from 0
to $\infty$. The quality factor (Q) of the Lorentzian feature is defined
as $\nu_{L}/FWHM$. If BLN centroid frequency is $>$ 10 Hz, then we call
it high-frequency noise (HFN) otherwise we call it low frequency noise
(LFN). The narrow feature with $Q>3$ is called QPO.
Details of the power spectral components observed in the different
branches of the Z-track are summarized in  Table 6.
\begin{sidewaystable}
{\scriptsize
\caption{The best fit model parameters of {\bf Model 1}. $kT_{e}$ is the electron temperature, $kT_s$ is 
the seed photon temperature in keV, $\Gamma_C$ is the photon index and $N_{Comp}$ is the 
normalization of the {\it nthComp} component. $\tau$ is the optical depth and $y-par$ is 
the Comptonization parameter. $E_L$ is the line energy, $R_{in}$ is the inner disc radius 
and $N_L$ is the normalization of the $diskline$ component. $\Gamma$ is the photon index 
and $N_{PL}$ is the normalization of power-law. $N_{PL}$ is in units of 
photons/s/cm$^2$/keV at 1 keV. $N_L$ is the line flux in units of photons/s/cm$^2$. F-test Prob 
is the chance improvement probability for an inclusion of a power-law. $F_{Comp}$ is 
the $3-80$ keV {\it nthComp} unabsorbed flux in units of ergs/s/cm$^2$ and $F_{PL}$ is 
the unabsorbed power-law flux in $3-80$ keV band in units of ergs/s/cm$^2$.} 
 \begin{tabular}{cccccccccc}
\hline
\hline
Parameters & HB1        & HB2 &        NB1            & NB2         & NB3& FB1 & FB2& FB3 & FB4\\
\hline
$kT_e$ (keV)&    3.20$\pm$0.05 &3.18$\pm$0.02& 3.17$\pm$0.03 &3.18$\pm$0.06 & 3.16$\pm$0.06 &2.95$\pm$0.07 & 2.90$\pm$0.04&2.85$\pm$0.04 & 2.90$\pm$0.06\\    
$\Gamma_C$& 1.89$\pm$0.03 & 2.10$\pm$0.03 & 2.20$\pm$0.02 & 2.40$\pm$0.03 & 2.47$\pm$0.03&2.34$\pm$0.04 & 2.22$\pm$0.05 & 2.07$\pm$0.04 & 1.95$\pm$0.05\\
$kT_s$ (keV) &0.6(fixed)&0.66$\pm$0.02&0.67$\pm$0.02 & 0.71$\pm$0.02 & 0.74$\pm$0.02 & 0.77$\pm$0.02& 0.99$\pm$0.04 & 1.04$\pm$0.05 & 1.12$\pm$0.05\\
$N_{Comp}$&0.90$\pm$0.08 & 1.09$\pm$0.03 & 1.11$\pm$0.05 & 1.01$\pm$0.04 & 0.88$\pm$0.05 & 0.80$\pm$0.04 & 0.49 $\pm$0.05 & 0.50$\pm$0.04 & 0.45$\pm$0.03\\
$\tau$   &10.35$\pm$0.61 & 8.93$\pm$0.25 & 8.38$\pm$0.14 &  7.46$\pm$0.17&7.21$\pm$0.16&8.05$\pm$0.26 & 8.73$\pm$0.38 & 9.71$\pm$0.38 & 10.44$\pm$0.57 \\
$y-par$&2.68$\pm$0.31 & 1.98$\pm$0.11 & 1.75$\pm$0.06 & 1.38$\pm$0.06 & 1.28$\pm$0.05 & 1.49$\pm$0.09 &1.73$\pm$0.15 & 2.09$\pm$0.16 & 2.48$\pm$0.27\\
$R_W$ (km) & 70.3$\pm$2.4 & 71.7$\pm$4.3 & 71.4$\pm$6.3 & 63.8$\pm$5.4 & 57.2$\pm$4.6 & 51.6$\pm$5.3 & 30.8$\pm$2.4 & 28.9$\pm$2.7 & 24.8$\pm$2.2 \\
$E_L$ (keV) & 6.79$\pm$0.08 & 6.91$\pm$0.05 & 6.95$\pm$0.07 & 6.98$\pm$0.06 & 6.92$\pm$0.04 & 6.94$\pm$0.06 & 6.89$\pm$0.07 & 6.92$\pm$0.08 & 6.89$\pm$0.09\\
$R_{in}$ ($GM/c^2$) & {\bf $<33$}  & 26.25$^{+5.56}_{-11.92}$ & 19.76$^{+11.51}_{-7.51}$ & 16.90$^{+8.51}_{-5.21}$ & 18.22$^{+6.78}_{-4.67}$ & 21.91$^{+10.05}_{-6.88}$ & 26.18$^{+15.97}_{-7.22}$ & 22.51$^{+14.25}_{-9.21}$&27.01$^{18.27}_{-10.51}$\\
$N_{L}$ ($\times$ 10$^{-2}$) &1.89$^{+0.12}_{-0.6}$ &  1.37$^{+0.23}_{-0.13}$ & 1.43$^{+0.26}_{-0.17}$&1.76$^{+0.15}_{-0.17}$&1.95$^{+0.23}_{-0.16}$&2.16$^{+0.31}_{-0.22}$ & 2.36$^{+0.23}_{-0.35}$ & 2.54$\pm$0.31 & 2.29$\pm$0.20 \\
$EW$ (eV)& 171.91$^{+10.92}_{-54.25}$& 117.10$^{+19.66}_{-11.13}$ & 128.62$^{+23.38}_{-15.29}$ & 182.32$^{+15.54}_{-17.61}$ & 222.34$^{+26.23}_{-18.25}$ & 211.55$^{+30.37}_{-15.68}$ & 166.54$^{+16.23}_{-24.70}$ & 143.50$\pm$17.52 & 116.41$\pm$20.33\\
$\Gamma$& 2.95$\pm$0.02 & 2.95(fixed) & 2.95(fixed) & 2.95(fixed) & 2.95(fixed) & 2.95(fixed) & 3.41$\pm$0.05 & 3.39$\pm$0.11 & 3.41$\pm$0.08 \\
$N_{PL}$ & 4.65$^{+0.75}_{-1.55}$ & 1.47$\pm$0.16 & 0.67$\pm$0.51 & 0.67$\pm$0.49 & 0.79$\pm$0.44 & 1.08$\pm$0.41 & 10.81$^{+3.15}_{-2.12}$&12.35$^{+3.04}_{-1.39}$ & 15.50$^{1.94}_{-2.53}$\\
$F_{Comp}$ ($\times$ 10$^{-9}$) & 13.18$\pm$0.31 & 14.46$\pm$0.33 & 14.12$\pm$0.31 & 12.35$\pm$0.28 & 10.97$\pm$0.25 & 11.76$\pm$0.27 & 14.14$\pm$0.33 & 17.79$\pm$0.41 & 20.42$\pm$0.47\\
$F_{PL}$ ($\times$ 10$^{-9}$)& 2.63$^{+0.6}_{-0.5}$ & 0.93$^{+0.34}_{-0.24}$ & 0.49$^{+0.31}_{-0.23}$ & 0.41$^{+0.30}_{-0.21}$ & 0.44$^{+0.21}_{-0.24}$ & 0.59$^{+0.36}_{-0.27}$ & 2.58$^{+0.31}_{-0.26}$ & 3.09$^{+0.37}_{-0.33}$ & 3.70$^{+0.35}_{-0.32}$\\
$F_{tot}$  ($\times 10^{-9}$) &  15.82$\pm$0.8 & 15.38$\pm$0.51 & 14.59$\pm$0.42 & 12.71$\pm$0.48 & 11.41$\pm$0.34 & 12.33$\pm$0.44 & 16.66 $\pm$0.45 & 20.8$\pm$0.55 & 24.15$\pm$0.58 \\
$F_{PL}$ (\%) &16.6 & 6.08 & 3.36 & 3.20 & 3.82 & 4.77 & 15.48 & 14.80 & 15.32 \\
$\chi^2/dof$ &   67/79 & 65/79&78/79 &90/79 & 88/79 & 101/79 & 85/78 & 78/78 & 67/78\\
$F-test Prob.$    &4.4e-9 &1.3e-5 & 2.7e-02 & 6.2e-02 & 6.6e-02 & 3.4e-02 &3.5e-09 & 6.8e-10 &2.2e-14\\
\hline
\hline
\end{tabular}
}
\end{sidewaystable}

\begin{sidewaystable}
{\scriptsize
\caption{The best fit spectral parameters of {\bf Model 1} with a Gaussian instead of a diskline for the iron line. $E_{Fe}$ is line energy for Gaussian line in units of keV, $\sigma_{Fe}$ is the width of line in units of keV and $N_{Fe}$ is the normalization of the line in units of $photons/s/cm^2$. Other paramters are same as in Table 1.}
\begin{tabular}{cccccccccc}
\hline
\hline
Parameters & HB1        & HB2 &        NB1            & NB2         & NB3& FB1 & FB2& FB3 & FB4\\
\hline
\hline
$\Gamma$ & 1.84$\pm$0.05 & 2.06$\pm$0.02 & 2.14$\pm$0.03&2.32$\pm$0.04&2.35$\pm$0.05 & 2.20$\pm$0.02 & 2.02$\pm$0.06& 1.91$\pm$0.07 & 1.85$\pm$0.07\\
$kT_e$ & 3.15$\pm$0.07& 3.12$\pm$0.05 & 3.12$\pm$0.06 & 3.13$\pm$0.08 & 3.02$\pm$0.07 & 2.83$\pm$0.05& 2.73$\pm$0.07&2.72$\pm$0.06&2.81$\pm$0.06\\
$kT_s$ & 0.6(fixed) & 0.59$\pm$0.04 & 0.61$\pm$0.04& 0.64$\pm$0.04 &0.64$\pm$0.05 &0.63$\pm$0.06 & 0.83$\pm$0.12& 0.87$\pm$0.12 & 1.02$\pm$.09\\
$\Gamma_{PL}$ & 3.06$\pm$0.06 & 2.95(fixed) & 2.95(fixed) & 2.95(fixed) & 2.95(fixed) & 2.95(fixed) & 3.25$\pm$0.19 & 3.22$\pm$0.18 & 3.35$\pm$0.10 \\ 
$N_{PL}$ & 6.95$\pm$1.2 & 1.81$\pm$ 0.51 & 0.80$\pm$0.52 & 0.63$\pm$0.48 & 0.95$\pm$0.45 & 1.19$\pm$0.42 & 8.15$\pm$3.2 & 8.67$\pm$3.3 & 14.75$\pm$3.2\\
$E_{Fe}$ (keV) & 6.4$\pm$0.49&6.57$\pm$0.21&6.52$\pm$0.18&6.45$\pm$0.21&6.43$\pm$0.35&6.48$\pm$0.17&6.61$\pm$0.18&6.69$\pm$0.28&6.54$\pm$0.29\\
$\sigma_{Fe}$ (keV) &0.99$\pm$0.24&0.92$\pm$0.20&1.05$\pm$0.22&1.09$\pm$0.16&1.04$\pm$0.27&1.06$\pm$0.2&0.95$\pm$0.18&0.95$\pm$0.29&1.20$\pm$0.22\\
$N_{Fe}$$(\times10^{-2})$& 2.82$\pm$0.81  &2.59$\pm$0.72&3.01$\pm$1.1&3.98$\pm$1.1&4.1$\pm$1.2&4.86$\pm$0.7&4.68$\pm$1.4&4.85$\pm$0.21&6.95$\pm$2.71\\
$\chi^2/dof$ &   66/79 & 61/79&75/79 &85/79 & 73/79 & 103/79 & 79/78 & 76/78 & 69/78\\
\hline
\hline
\end{tabular}
}
\end{sidewaystable}
\begin{sidewaystable}
{\scriptsize
\caption{Best fit parameters for {\bf Model 2}. $kT_{in}$ is the inner disk temperature, $N_{dbb}$ is the normalization of the diskbb, $R_{eff}$ is the effective radius of inner disk, $kT_{BB}$ is the blackbody temperature and $N_{BB}$ is the normalization of the blackbody component. F-test Prob. is the chance improvement probability for an inclusion of the power-law component.}
\begin{tabular}{cccccccccc}
\hline
\hline
Parameters & HB1 & HB2 & NB1 & NB2 & NB3 & FB1 & FB2 & FB3 & FB4\\
\hline
$kT_{in}$ (keV) & 1.73$\pm$0.05 &1.73$\pm$0.05 & 1.74$\pm$0.05 & 1.63$\pm$0.04 & 1.62$\pm$0.04 & 1.73$\pm$0.04 & 1.83$\pm$0.04&1.98$\pm$0.06 & 2.24$\pm$0.12\\
$N_{dbb}$ & 59.71$\pm$10.55 &  76.51$\pm$10.35 & 76.96$\pm$8.95 & 96.85$\pm$11.15 & 97.02$\pm$11.45 & 79.85$\pm$9.05 & 75.55$\pm$12.5 & 63.82$\pm$11.45 & 44.25$\pm$7.15\\
$R_{eff}$ (in km) & 32.65$\pm$2.83 &36.98$\pm$2.48 & 37.05$\pm$2.15 & 41.58$\pm$2.39 & 41.64$\pm$2.45 & 37.75$\pm$2.13 & 36.68$\pm$2.03 & 33.75$\pm$3.02&28.10$\pm$2.26\\
$kT_{BB}$ (keV) & 2.77$\pm$0.03 & 2.71$\pm$0.04 & 2.70$\pm$0.05 & 2.60$\pm$0.05&2.57$\pm$0.04 & 2.61$\pm$0.06 & 2.58$\pm$0.05 & 2.65$\pm$0.05 & 2.79$\pm$0.07\\ 
$N_{BB}$ & 10.19$\pm$0.85 & 10.21$\pm$0.98 & 8.65$\pm$0.95 & 7.61$\pm$0.95 & 6.75$\pm$0.93&7.19$\pm$1.23 & 11.38$\pm$1.92&13.32$\pm$2.05 & 12.06$\pm$3.55 \\
$R_{BB}$ (km) & 4.2$\pm$0.1 & 4.2$\pm$0.1 & 3.8$\pm$0.2 & 3.6$\pm$0.2 & 3.4$\pm$0.2 & 3.5$\pm$0.3 & 4.4$\pm$0.3 & 4.7$\pm$0.3 & 4.5$\pm$0.6\\
$R_{Beff}$ (km) & 9.3$\pm$0.3 & 9.3$\pm$0.2 & 8.55$\pm$0.5 & 8.05$\pm$0.5 & 7.5$\pm$0.5 & 7.8$\pm$0.7 & 9.8$\pm$0.8 & 10.6$\pm$0.7 & 10.2$\pm$1.1\\
$\Gamma_{PL}$ & 2.90$\pm$0.07 & 3.11$\pm$0.15 & 3.33$\pm$0.17 & 3.25$\pm$0.23 & 3.18$\pm$0.32 & 3.29$\pm$0.36 & 2.87$\pm$0.25 & 2.93$\pm$0.25 & 3.09$\pm$ 0.14\\
$N_{PL}$ & 5.22$\pm$1.17 & 4.86$\pm$1.44 & 5.33$\pm$1.68 & 3.52$\pm$1.95 & 2.49$\pm$1.85 & 2.56$\pm$2.49 & 1.96$\pm$1.08 & 2.64$\pm$1.48 & 3.09$\pm$4.78\\
$E_{L}$ & 6.87$\pm$0.87 & 6.89$\pm$0.11 & 6.86$\pm$0.12 & 6.93$\pm$0.08 & 6.87$\pm$0.06 & 6.86$\pm$0.06 &  6.89$\pm$0.04 & 6.91$\pm$0.07 & 6.89$\pm$0.07 \\
$R_{in} (GM/c^2)$ &39.85$^{+60}_{-20}$ & 40.95$^{+75}_{-22}$ & 39.5$^{+60}_{-21}$ & 21.6$^{+17}_{-12}$ & 20.78$^{+15}_{-9.8}$ & 24.56$^{+24}_{-9.8}$ & 24.75$^{+22}_{-9.8}$ & 20.81$^{+16}_{-9.4}$& 21.37$^{+20}_{-9.4}$\\
$N_{L} (\times 10^{-2}$) &1.17$\pm$0.14 & 1.09$\pm$0.15 & 1.04$\pm$0.10 & 1.44$\pm$0.20 & 1.69$\pm$0.21 & 1.99$\pm$0.27 & 2.41$\pm$0.27&2.69$\pm$0.37 & 2.61$\pm$0.28\\
$F_{total} (\times 10^{-9})$ &  15.71$\pm$0.39 & 15.40$\pm$0.43 &14.29$\pm$0.43&12.25$\pm$0.38 & 11.17$\pm$0.41 & 12.32$\pm$0.41 & 16.82$\pm$0.72 & 20.8$\pm$0.95 & 24.31$\pm$0.93\\
$F_{PL} (\times 10{-^9})$   &3.23$\pm$0.23 & 2.02$\pm$0.29 & 1.36$\pm$0.32 & 1.12$\pm$0.29 & 0.90$\pm$0.34 & 0.75$\pm$0.33 & 1.35$\pm$0.38 & 1.51$\pm$0.39 & 2.09$\pm$0.30\\
$F_{PL}$(in \%) &20.59 & 13.10 & 9.51 & 9.12 & 8.05 & 6.08& 8.02 & 7.25 & 8.60 \\
$\chi^2/dof$   & 63/78 & 62/78 & 77/78 & 81/78 & 81/78  &113/78 &85/78 &80/78 &70/78\\
$F-test$ Prob. & 4.7e-22 & 1.2e-14 & 9.1e-7 & 1.8e-4 & 3.0e-3 & 6.9e-2 & 7.5e-6 & 1.1e-5 & 2.2e-8\\ 
\hline
\end{tabular}
}
\end{sidewaystable}
\begin{sidewaystable}
{\scriptsize
\caption{Best fit parameters for {\bf Model 3}. $f_{refl}$ is reflection fraction, $A_{Fe}$ is iron abundance and $\log{\xi}$ is the logarithm of ionization parameter.}
\begin{tabular}{cccccccccc}
\hline
\hline
Parameters & HB1           & HB2 & NB1 & NB2 & NB3 & FB1 & FB2 & FB3 & FB4\\
\hline

$f_{refl}$ & 0.29$\pm$0.11 & 0.24$\pm$0.09&0.23$\pm$0.04&0.36$\pm$0.04&0.45$\pm$0.07&0.51$\pm$0.07&0.37$\pm$0.09&0.27$\pm$0.09&0.23$\pm$0.08\\
$A_{Fe}$    &0.49$\pm$0.23& 0.5(fixed)   &0.5(fixed) &0.5(fixed) & 0.5(fixed)&0.49$\pm$0.31&0.5(fixed) & 0.5(fixed)&0.5(fixed)\\
$\log{\xi}$ &2.83$\pm$0.13& 3.30$\pm$0.12 &3.02$\pm$0.21&3.06$\pm$0.31&3.01$\pm$0.4&2.99$\pm$0.31&3.01$\pm$0.22&2.99$\pm$0.38&2.98$\pm$0.55\\
$\Gamma_C$ & 1.89$\pm$0.03& 2.02$\pm$0.03 &2.15$\pm$0.04&2.27$\pm$0.05&2.36$\pm$0.04&2.23$\pm$0.03&2.22$\pm$0.09&1.95$\pm$0.11&1.82$\pm$0.07\\
$kT_e$ (keV)&3.10$\pm$0.08&3.02$\pm$0.04  &3.06$\pm$0.06&2.89$\pm$0.06&2.85$\pm$0.05&2.70$\pm$0.09&2.79$\pm$0.08&2.68$\pm$0.06&2.75$\pm$0.04\\
$kT_s$ (keV)&0.6(fixed)& 0.59$\pm$0.03    &0.72$\pm$0.05&0.68$\pm$0.03&0.71$\pm$0.03&0.74$\pm$0.05&1.06$\pm$0.05&0.93$\pm$0.08&0.95$\pm$0.09\\
$\Gamma_{PL}$ & 2.96$\pm$0.06&2.95(fixed) &2.95(fixed)&2.95(fixed)&2.95(fixed)&2.95(fixed) &3.47$\pm$0.09&3.20$\pm$0.12&3.14$\pm$0.12\\
$N_{PL}$ &     5.60$\pm$0.86& 2.56$\pm$0.4&1.30$\pm$0.41&1.49$\pm$0.1&1.46$\pm$0.25&1.65$\pm$0.36&15.88$\pm$2.77&8.48$\pm$3.1&8.07$\pm$2.9\\
$\chi^2/dof$ & 56/79 & 61/80 & 72/80 & 78/80 &68/80 &102/79 &72/79 &81/79&74/79\\
\hline
\hline
\end{tabular}
}
\end{sidewaystable}
\begin{sidewaystable}
{\scriptsize
\caption{Best fit parameters for {\bf Model 4}. $f_{refl}$ is reflection fraction, iron abundance is fixed 1.0.  $\log{\xi}$ is the logarithm of ionization parameter. $kT_{BB}$ and $kT_{in}$ is the blackbody and inner disk temperature (in keV) respectively. The parameter norm is BBREFL normalization. $\Gamma$ and $N_{PL}$ are index and normalization of the power-law component. }
\begin{tabular}{cccccccccc}
\hline
\hline
Parameters & HB1           & HB2 & NB1 & NB2 & NB3 & FB1 & FB2 & FB3 & FB4\\
\hline
$R_{in}$(ISCO)& 14.02$\pm$5.4& 18.5$\pm$4.2 & 18.6$\pm$3.9 & 15.4$\pm$4.3& 13.3$\pm$4.1& 14.4$\pm$3.9 & 13.2$\pm$3.7 &12.35$\pm$4.8 & 12.45$\pm$4.4\\
$\log\xi$ &2.64$\pm$0.5 & 2.66$\pm$0.4 & 2.68$\pm$0.45& 2.81$\pm$0.4 & {\bf 2.72$\pm$0.2} & 2.71$\pm$0.3 & 2.73$\pm$0.4 &2.64$\pm$0.3 & 2.73$\pm$0.3\\   
$f_{refl}$ & 0.36$\pm$0.14 & 0.34$\pm$0.12&  0.35$\pm$0.12& 0.61$\pm$0.06 & 0.88$\pm$0.11&0.86$\pm$0.10&0.76$\pm$0.22 & 0.72$\pm$0.32 & 0.67$\pm$0.45\\
$norm(1 \times 10^{-26}$) & 17.5$\pm$7.6 & 15.9$\pm$4.4& 10.6$\pm$5.2 & 6.8$\pm$1.9 & 6.4$\pm$2.5 & 7.23$\pm$3.5 & 10.8$\pm$4.5 & 17.4$\pm$5.5& 14.2$\pm$4.9\\
$kT_{BB}(keV) $ &2.65$\pm$0.03& 2.54$\pm$0.03& 2.50$\pm$0.04& 2.39$\pm$0.02& 2.35$\pm$0.04&2.35$\pm$0.05 & 2.42$\pm$0.06 & 2.52$\pm$0.15 & 2.69$\pm$0.22\\
$kT_{in} (keV) $&1.64$\pm$0.10 & 1.60$\pm$0.05 & 1.59$\pm$0.05&  {\bf 1.50$\pm$0.04}& 1.51$\pm$0.06& 1.58$\pm$0.06 & 1.77$\pm$0.07& 1.99$\pm$0.25 & 2.29$\pm$0.5 \\
$N_{dbb}$&70.5$\pm$20.5 & 104.4$\pm$13.6 & 111.6$\pm$12.2 & 127.8$\pm$14.3& 122.7$\pm$12.4& 104.3$\pm$15.5 & 76.5$\pm$19.4 & 56.8$\pm$25.3& 36.7$\pm$ 17.9\\ 
$\Gamma$& 2.85$\pm$0.06 & 2.90(fix) & 2.90(fix) & 2.90(fix) & 2.90(fix) & 2.90(fix) & 3.01$\pm$0.15 & 3.04$\pm$0.18 & 3.15$\pm$0.13\\
$N_{PL}$ & 5.41$\pm$0.91 & 2.93$\pm$ 0.45 & 2.03$\pm$0.35 & 2.05$\pm$0.36 & 1.90$\pm$0.34 & 1.91$\pm$0.38 & 3.99$\pm$0.85 & 4.76$\pm$1.25&7.55$\pm$1.32\\ 
$\chi^2/dof$ & 57/78 & 60/79 & 73/79 & 67/79& 59/79 & 95/79 & 65/78& 68/78 & 56/78\\
\hline
\end{tabular}
}
\end{sidewaystable}
\begin{table}
\caption{Model components of PDS in the frequency range $0.07-100$ Hz. Three noise features 
LFN (Low-frequency noise), HFN (High-frequency noise) and VLFN (Very low-frequency noise) 
are observed (in the bracket). The QPO is observed in the NB2 (in the bracket).}

\begin{tabular}{cc}\hline
Sections & PDS Model Components\\
\hline
HB1      & Lorentzian (LFN) + Lorentzian (HFN) \\
HB2      & Lorentzian (LFN) \\
NB1      & Lorentzian (LFN) \\
NB2      & Power-law (VLFN) + Lorentzian (QPO)\\
NB3      &Power-law (VLFN) + Lorentzian (LFN)\\
FB1      &Lorentzian (LFN)\\
FB2      &Power-law (VLFN) + Lorentzian (LFN)\\
FB3      &Power-law (VLFN) \\
FB4      &Power-law (VLFN) \\
\hline
\hline
\end{tabular}
\end{table}
\begin{sidewaystable}
{\scriptsize
\caption{The best fit results obtained by fitting the PDS at different parts of the Z-track. 
$\nu_L$ is the centroid frequency, $FWHM$ is the full-width-half-maximum, $LN$ is the 
normalization of Lorentzian feature. $rms$ is the integrated rms in the $0.07-100$ Hz. 
$\alpha$ is the index and $A$ is the normalization of the power-law component.}
\begin{tabular}{cccccccccc}
\hline
Parameters & HB1 & HB2 & NB1 & NB2 & NB3 & FB1 & FB2 & FB3 & FB4 \\
\hline
\hline
& & & & LFN (Lorentzian) & & & & & \\
\hline
$\nu_L (Hz)$    &2.53$\pm$0.26 & 0.0(fixed) &3.32$\pm$1.07 &7.42$\pm$0.23 &8.09$\pm$0.38 &6.25$\pm$4.16 &7.88$\pm$1.58 & & \\
$FWHM (Hz) $   &7.20$\pm$1.05 & 9.16$\pm$0.95 &8.96$\pm$4.59 &1.52$\pm$0.59 &7.13$\pm$1.12 &31.02$\pm$9.72 &12.29$\pm$3.47 & & \\
$LN$ $(\times 10^{5})$  &20.42$\pm$1.32&  13.93$\pm$1.25 &2.56$\pm$0.68 &8.47$\pm$1.89&13.12$\pm$1.24 & 3.75$\pm$0.78&3.22$\pm$0.56&&\\
$rms (\%) $ & 3.96$\pm$0.46 & 3.11$\pm$0.44 &  1.58$\pm$0.65&1.41$\pm$0.29&3.54$\pm$0.37 & 3.22$\pm$1.06 &  2.18$\pm$0.47& & \\
\hline
& & & & HFN (Lorentzian) & & & & & \\
\hline
$\nu_L (Hz) $    &25.68$\pm$3.52 & & & & & & & & \\
$FWHM (Hz) $   &22.44$\pm$7.33 & & & & & & & \\
$LN$ $(\times 10^{5})$      &2.34$\pm$0.61 & & && &&&&\\
$rms (\%) $ & 2.59$\pm$0.65 & & & & & & & & \\
\hline
& & & & VLFN (Power-law) & & & & & \\
\hline
$\alpha$ & & & &-0.48$\pm$0.21 &-1.01$\pm$0.22 & &-1.81$\pm$0.26 &-1.55$\pm$0.25 &-1.45$\pm$0.29 \\
$A$($\times 10^{5}$) & & & &1.20$\pm$0.69 &2.55$\pm$1.33 & &0.63$\pm$0.24 &1.77$\pm$1.12 &3.55$\pm$0.95\\
$rms (\%) $ & & & &1.59$\pm$1.21 & 1.39$\pm$0.24 & &0.94$\pm$0.01 & 1.29$\pm$0.018 &1.72$\pm$0.05 \\
\hline
Total-rms(\%) &4.73$\pm$0.70 & 3.11$\pm$0.44 & 1.58$\pm$0.65 & 2.12$\pm$1.25 & 3.80$\pm$0.35 & 3.22$\pm$1.06 & 2.38$\pm$0.47 & 1.32$\pm$0.025 & 1.72$\pm$0.05 \\
\hline
$\chi^2$/dof & 26/54 & 32/57 &56/57 &26/55 & 42/55 &60/57 &40/55 &33/58 & 44/58\\
\hline
\end{tabular}
}
\end{sidewaystable}
\begin{figure}[h]
\includegraphics[width=0.32\textwidth,angle=-90]{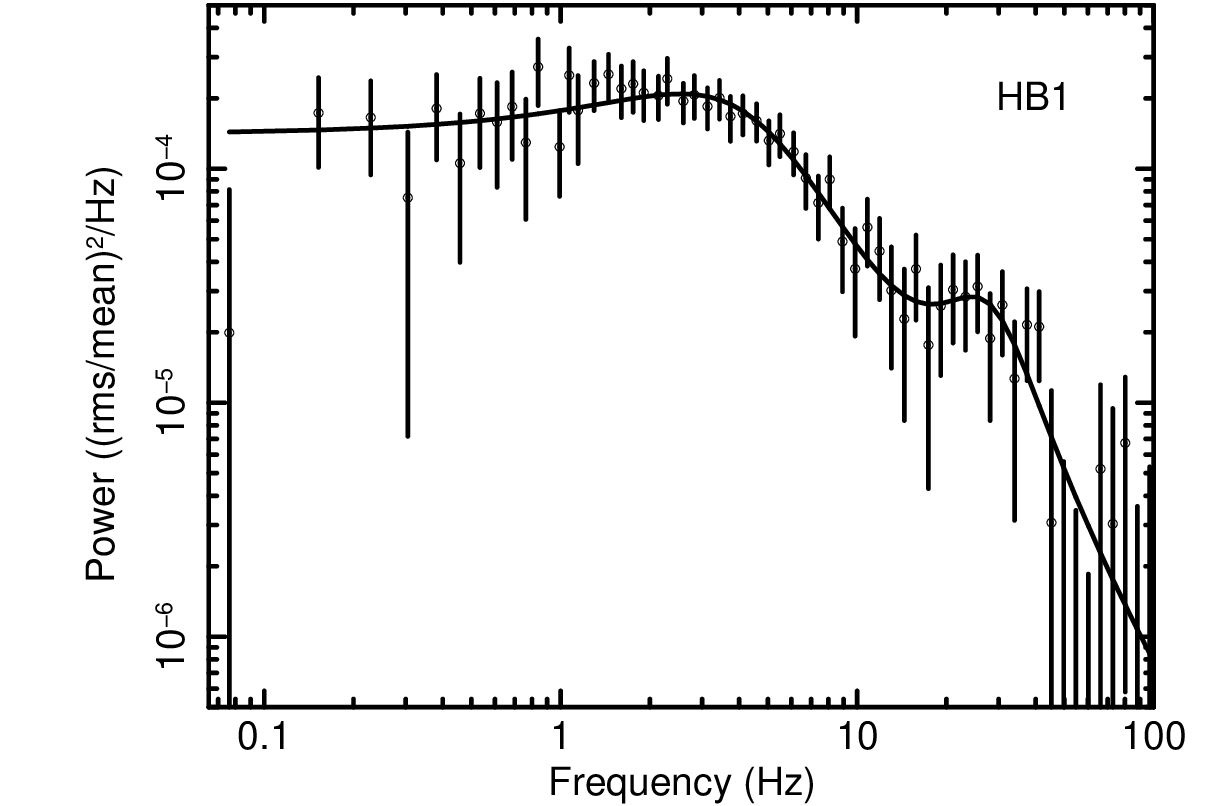}
\includegraphics[width=0.32\textwidth,angle=-90]{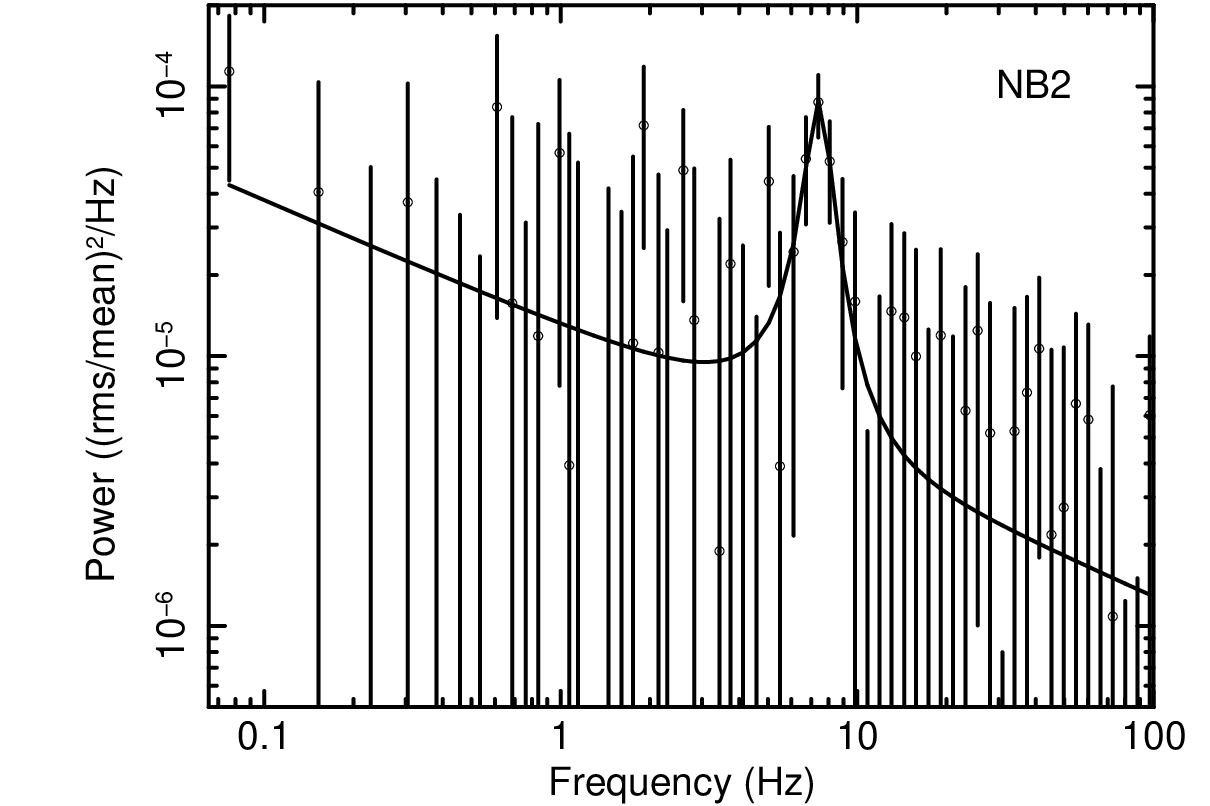}
\includegraphics[width=0.32\textwidth,angle=-90]{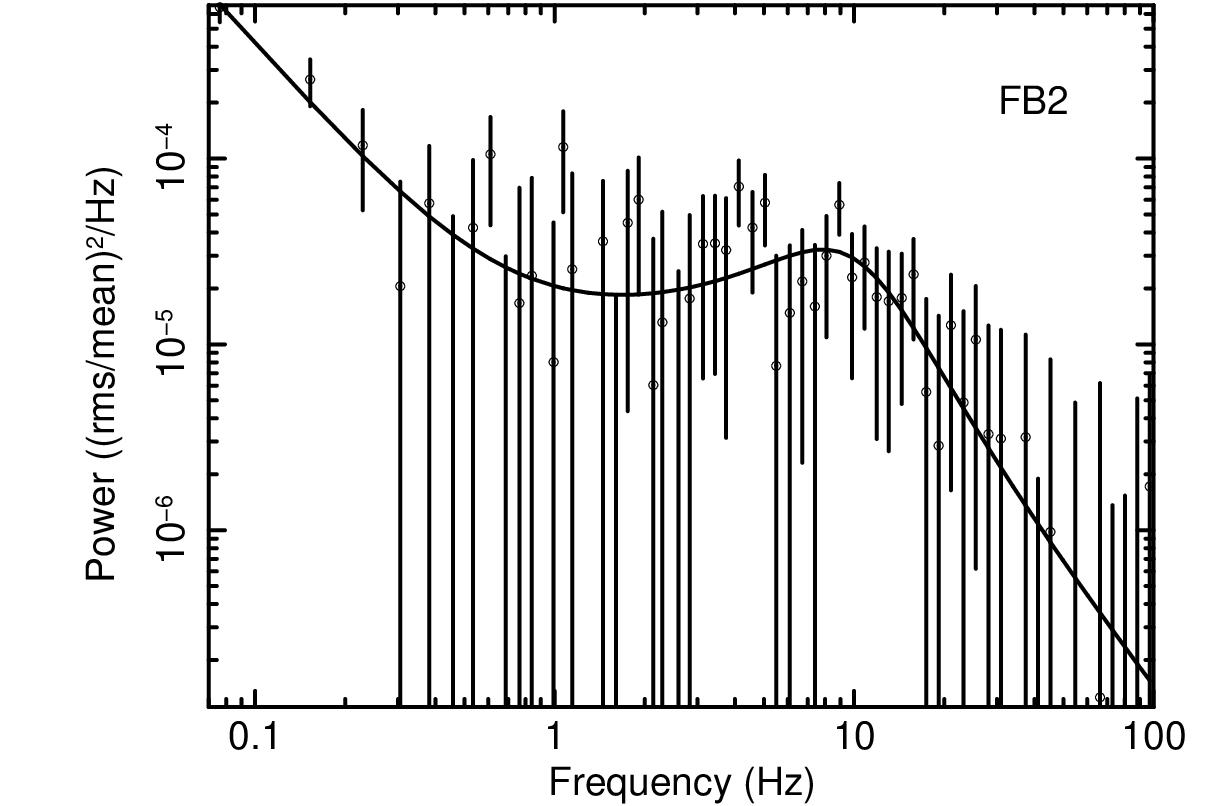}
\caption{PDS and the best fit model at three different parts of the Z-track. HB1: top-left panel, NB2: top-right panel and FB2: bottom panel. The used model is described} in Table 6 and best fit parameters are given in Table 7.  A significant QPO is clearly seen in the PDS of NB2.
\end{figure}

\section{Results}

\subsection{Spectral Properties} 

The $3.0-80.0$ keV X-ray spectra extracted from the different sections of
the Z-track can be fitted by two different phenomenological models. {\bf
Model 1} consists of a Comptonized emission, a relativistic iron line
feature and a high energy tail. Table 1 shows the best fit parameters. The
F-test chance improvement probability for an inclusion of a power-law
tail is also listed in the Table 1. {\bf Model 2}, which consists of
disc emission, a blackbody emission component, a relativistic iron
line and a non-thermal tail, is also statistically good description
of the data. In Figure 3, we show the unfolded spectra for HB1, NB2 and
FB2 fitted with these two models. The best fit parameters of {\bf Model 2} are shown in the Table 3. We also note that {\bf Model 3} and {\bf Model 4}
which includes reflection component to explain the origin of iron line
is also a good description of the spectra for all the sections of the
Z-track. In Figure 4a we show the unfolded spectrum of the HB1 section fitted
with {\bf Model 3} and in Figure 4b we show that fitted with {\bf Model 4}. 
In  Figure 5a, 5b and 5c, we show the variation of the best
fit spectral parameters for {\bf Model 1} (left panel), {\bf Model 2}
(middle panel)  and {\bf Model 3} (right panel) respectively. The best fit parameters of {\bf Model 3} and {\bf Model 4} are shown in Table 4 and 5 respectively.
 
The contribution of the high energy tail to the total $3.0-80.0$ keV
flux is the highest in the HB1 (see Table 1 and Table 3). It gradually
fades as the source moves down the HB and then down the NB. As the source
moves up along the FB the power-law flux systematically increases. The
two different phenomenological models used here give different values of
the power-law flux and the parameters of the power-law. However, trend
in the variation of the power-law strength along the Z-track is similar.

Modelling the iron line feature with  $diskline$ model allows to constrain
the  inner radius of the disc. Since error present in the inferred
inner disc radius is large, we can not comment much about its evolution along
the Z-track. The inner disc radius obtained from {\bf Model 1} varies
between $\sim$ $10.0-45.0$ $R_g$ being lowest at the upper HB. However,
$R_{in}$ inferred from the {\bf Model 2} is in the range of $\sim$
$20.0-40.0$. The equivalent width of the iron line is found to be in the
range of $\sim$ $115-220$ eV.
The dominant emission component of {\bf Model 1} present in the $3-80$
keV spectra is thermal Comptonized emission, modelled with {\it nthComp}
model of XSPEC.  The plasma temperature obtained by fitting with this
model is $\sim$ 3.20 keV in the HB and NB, and $\sim$ 2.90 keV in
the FB. The photon index ($\Gamma_C$) of Comptonized emission shows
systematic evolution along the Z-track (see Figure 5a). The index
$\Gamma_C$ initially increases as the source moves along the HB and
then down the NB. $\Gamma_C$ again decreases as the source moves up
along the FB. Figure 5a shows the evolution of the best fit spectral
parameters as a function of the position on the HID. The optical depth
$\tau$ is calculated using formula given in \cite{agrawal18} and the
Comptonization parameter $y$ is calculated using the formula,

\begin{equation}
y= \frac{4kT_e}{m_e c^2} max(\tau, \tau^2),
\end{equation}
where $kT_e$ is the electron temperature, $m_e$ is the mass of the
electron, $\tau$ is the optical depth and c is the speed of light.
The seed photon radius $R_W$ of the spherical area emitting the blackbody spectrum is given by \citep{jjm},
\begin{equation}
R_W = 3 \times 10^4 D\sqrt{\frac{f_{bol}}{1+y}}/(kT_s)^2,
\end{equation}

where $kT_s$ is the seed photon temperature in keV, $D$ is the distance to the source in kpc and $f_{bol}$ is the bolometric ($0.1-100$ keV) luminosity of the Comptonized component. The seed photon radius $R_W$ varies from 70 km to 25 km as the source moves along the Z-track from the HB to FB, suggesting that inner disc approaches the central source as the source moves along the Z-track from the HB to FB through NB. This is expected if mass accretion rate is increasing from the HB to FB.

The optical depth $\tau$ of the corona decreases as the source moves
down from the HB to the lowest part of the NB. It again increases as
the source moves up along the FB. The Comptonization parameter ($y-par$)
decreases as the source moves down the HB and then again towards the
lower part of the NB. The increase in $y-par$ is observed as the source
moves up the FB (see Figure 5a and Table 1).

Fitting the data with {\bf Model 2} gives inner-disc temperature in the
range of $1.6-2.2$ keV (see Table 3). In the diskbb model, inner disk
radius $R_{d}$ is given  by $\sqrt{N_{dbb}/\cos\theta} D_{10}$, where
$D_{10}$ is distance to the source in 10 kpc and $\theta$ is the
inclination angle of the disk. \cite{ST95} argue that if the electron
scattering dominates the opacity, the local spectrum will significantly
deviate from the blackbody and is approximated by a diluted blackbody. The
local specific flux for diluted blackbody is given by,
\begin{equation}
F_\nu = \frac{\pi}{f_{col}^4} B_\nu(f_{col} T_{eff}),
\end{equation}
where $f_{col}$ is the spectral hardening factor, $T_{eff}$ is the
effective temperature and $B_\nu$ is the Planck function.  \cite{Davis05}
studied the dependence of $f_{col}$ on the luminosities and inclination
angle. For an inclination angle of $i=45^\circ$ and $L/L_{Edd}$ = 0.3
they found $f_{col}=1.6$. The effective inner disc radius and effective
temperature is given by, 
\begin{equation} 
R_{eff} = f_{col}^2 R_d, \quad T_{eff} = kT_{in}/f_{col}. 
\end{equation} The estimated effective radius
varies between $\sim$ $28-42$ km for an inclination angle $i=40^\circ$ and
distance to the source  D = 12.6 kpc \citep{Kuulkers03}. $R_{eff}$ increases as the source moves from HB1 to NB3 and then decreases as it moves along the FB. The trend in the variation of $R_{eff}$ and $R_W$ is similar along the FB. However, it is opposite in the HB and NB. The effective
temperature is found to be in the range of $1.0-1.4$ keV. The blackbody
component has temperature in the range of $2.6-2.8$ keV. 
The blackbody radius in {\it bbodyrad} model is given by,
\begin{equation}
R_{BB}  = \sqrt{N_{BB}} D_{10},
\end{equation}

where $R_{BB}$ radius of blackbody emitting surface in units of km,
$N_{BB}$ is normalization of {\it bbodyrad} and $D_{10}$ the source
distance in the units of 10 kpc. The blackbody radius is found to be
$\sim$ 4 km at different parts of the Z-track.

\cite{Foster86} fitted a series of numerical models assuming constant density atmosphere to the spectra of type I X-ray bursts observed by {\it EXOSAT}. They derived relationship between the effective blackbody temperature  and the fitted or colour temperature. They found colour correction factor ($f_{col}$) in the range of $\sim 1.3-1.6$. Using the average colour correction factor $f_{col}$ $\sim$ 1.5, we compute the effective blackbody radius $R_{Beff}$ using equation,
\begin{equation}
R_{Beff}  = f_{col}^2 R_{BB},
\end{equation}
The effective blackbody radius is found to be in the range of $8-10$ km (see Table 3).

{\bf Model 3} gives iron abundance $A_{Fe}$ $\sim$ 0.5. The reflection
fraction $f_{refl}$ varies between $0.2-0.5$ along the Z-track and logarithm
of ionization parameter is found to be around 3.0 on the Z-track (see
Table 4 and Figure 5c). The X-ray spectrum of the section HB1,
fitted with {\bf Model 3} is shown in the Figure 4a.  We note that the
iron line observed in the spectra of the source at different parts of
the Z-track is narrow  and does not require an additional relativistic
smearing. We checked this by introducing additional bluring using the
convolution model {\it rdblur} and found that the fit is not improved. We
also plot the equivalent width of iron line derived using {\bf Model
1} against reflection fraction derived using {\bf Model 3}. A clear
correlation between these two parameters can be seen (see Figure 6).
{\bf Model 4} gives reflection fraction $f_{refl}$ in the range of
$0.3-0.9$.  It also suggests that  the iron line is narrow and disc is
truncated away from the central compact object ($10 - 22$ $R_{ISCO}$). The logrithm of ionization parameter is found to be around $\sim 2.5-3.0$.  In {\bf Model 4}, we have fixed the iron abundance at 1.0.

\subsection{Timing Behaviour}

The timing analysis of the source revealed the presence of a QPO in the
middle NB (NB2) at the frequency 7.42$\pm$0.23 Hz (see the top-right panel
of Figure 7). The F-test probability that the fit is improved by chance for
an inclusion of this narrow feature is $\sim$ 3.1e-7 and the significance
of this QPO is 5.1$\sigma$. The quality factor of the NBO is found to
be $\sim$ 4.88 and rms is found to be 1.41$\pm$0.29\%. We also detect
different types of band limited and red noise in the PDS. The types
of the noise component detected in the different branches are listed
in Table 6 and their best fit parameters along with rms are given in
 Table 7. In the top left HB (HB1), we find both LFN and HFN. In the HB2,
we find only LFN. In the NB1, we detect only LFN. But in NB2 and NB3, we
detect both VLFN and LFN. Again, in the FB1 we detect only LFN. In the
FB2, we detect both LFN and VLFN. The FB3 and FB4 PDS show the presence
of only VLFN feature. In Figure 7, we have shown three representative
PDS at different sections of the Z-track. The integrated rms in the frequency range 
$0.07-100$ Hz is found to be in the range of $1.3-4.75$\%.

\section{Discussion}

In this paper, we report the results obtained using 50 ks LAXPC
observations of the Z-source GX 17+2. In the HID, the source displayed
all the three branches of the Z-diagram making it possible to study the
broadband spectral evolution of the source in $3.0-80.0$ keV energy band
along the complete Z-track.  

{\bf Model 1} requires a strong hard X-ray tail ($>$ 30
keV) contributing 17\% of the total $3.0-80.0$ keV flux in the upper HB (HB1). This hard
X-ray tail can be described by a power-law model with photon index
$\Gamma$=2.95$\pm$0.02. The hard X-ray tail becomes weaker as the source
moves down the HB.  In the HB2 the contribution of the hard tail is $\sim$
6.1\%.  As the source moves further down the NB the hard tail becomes
much weaker contributing only $3-4$ \% of the total flux. 
{\bf Model 2} has three components: emission from the thermal disk,
blackbody emission from the boundary layer or NS surface and a
power-law emission from a corona.  This model is similar to that used
by \cite{Lin12}. Note that \cite{Lin12} used cutoff power-law (CPL)
with cutoff at around 10 keV to model the excess emission above 20 keV.
However, we find a strong power-law emission extending above 30 keV without showing any signature of a cutoff.  The power-law tail contributes
$\sim$ 20\% of the total flux in the upper HB when the spectra are fitted
with {\bf Model 2}. The power-law component diminishes as the source
moves down the HB and further down  in the NB.  Though {\bf Model 2} predicts
a significant power-law emission in the NB, its contribution above 40
keV is almost negligible. This is due to fact that power-law component
is steep in this region of the Z-track.  The power-law again becomes
stronger as the source moves up the FB.  The effective inner disc radius
$R_{eff}$ estimated using {\bf Model 2} increases as the source moves
from the HB to the lower NB. The radius $R_{eff}$ decreases as the source
moves up the FB (see Figure 4b). A similar variation in the inner disc
radius has been seen during the {\it RXTE} observations \citep{Lin12}.
The radius of neutron-star/boundary-layer obtained using {\bf Model 2}
is only $\sim$ 4 km. This radius has been obtained assuming isotropic emission from the neutron star surface. However, it is possible that the emission from the neutron star surface is anisotropic. The radius obtained assuming isotropic emission will be smaller compared to that obtained assuming anisotropic emission. We also corrected the radius for scattering in the the neutron star atmposphere assuming constant density atmosphere \citep{Foster86}. In this case, the corrected radius is close to the radius of the neutron star ($8-10$ km).

It is worth mentioning that the contribution of the power-law component is found to be model dependent. However, both models ({\bf Model 1} and {\bf Model 2}) predict the presence of a strong power-law component extending above 30 keV in the FB.

Previous {\it BeppoSAX} observations of the source have also shown the
presence of a strong hard X-ray tail in the upper HB which gradually
faded as the source moved down the HB and completely vanished in the
NB \citep{disalvo00}.  An another {\it BeppoSAX} observation of this
source where the source displayed only  HB \citep{Farinelli05} revealed
a hard X-ray tail with photon index $\sim$ 2.8 in the upper HB. The hard
tail disappeared as the source moved to the lower HB \citep{Farinelli05}.
More importantly, we also find the evidence of strong hard X-ray
tails in the middle and upper FB. The power-law tail is steep
having photon index of $\sim$ $3.3-3.4$. The hard X-ray tail observed in
this section of the FB contributes $\sim$ 15\% of the total $3.0-80.0$
keV flux. Previously, hard X-ray tail has been detected in the FB
of this source using HEXTE (High Energy X-ray Timing Experiment) data
\citep{Ding15}. Other sources which have shown hard tail in the FB are
Sco X-1 and GX 5-1 \citep{asai94, damico1}. A hard X-ray tail with a
 flat or inverted index was reported in the FB of  Sco X-1 \citep{damico1,dai07}
with {\it HEXTE/RXTE}. However, {\it INTEGRAL} observation of Sco X-1 revealed
that at the top of the FB the hard X-ray tail disappeared \citep{revni14,disalvo06}, suggesting transient nature of the hard X-ray tail.

GX 5-1 has also shown hard X-ray tails in the FB, however this component
was fainter in the FB compared to the NB \citep{asai94}. In GX 349+2, the
hard X-ray tail with photon index $\sim$ 1.9 was found in the non-flaring
state corresponding to the lower part of the NB \citep{disalvo01} and
disappeared in the flaring branch. Similarly, 2001 BeppoSAX observations of GX 349+2 when the source was probably in the FB suggested the absence of the hard X-ray tail in the X-ray spectra of the source \citep{Iaria04}.

Simultaneous X-ray and radio observations of GX 17+2 indicate
that the radio flux is correlated with the position on the HID
\citep{penninx88,mig07}. The radio flux decreases from the HB to the
NB and is quenched in the FB \citep{mig07}. The presence of the hard
X-ray tail in the HB suggests that most probably
Comptonization of the seed photons in a relativistic jet produces
the hard X-ray tail \citep{disalvo00,disalvo02}. However, present
detection of a strong hard tail in the FB of GX 17+2, where radio flux
disappears, can not be explained in this framework, a simultaneous radio/X-ray observation in this state would be needed to strengthen this statement. Other competing
model is up-scattering of the soft seed photons from the neutron
star surface by relativistic electrons in bulk motion of converging
flow \citep{Titarchuk98,Farinelli09}. However, in the FB radiation
pressure can slow down or can stop the bulk flow causing quenching of
hard X-ray tail \citep{Farinelli07}. This suggests that a significant
detection of hard X-ray tail in the FB can not be explained in the
frame work of bulk Comptonization. Hence, most probably the hard X-ray
tail is produced by hybrid thermal/non-thermal electrons in the corona
\citep{coppi99}. \cite{Farinelli05} used this model to explain the hard
tail observed in the HB of GX 17+2. The hybrid thermal/non-thermal model
was also applied to describe the hard excess seen in the spectra of Sco X-1 \citep{dai07}.

In {\bf Model 1} the emission from the source in the $3.0-80.0$
keV band is dominated by the contribution from an optically thick
($\tau \sim 7-10$) and cool ($kTe \sim 3~keV$) {\bf corona} which is in
accordance with the results obtained from the previous {\it BeppoSAX}
observations \citep{disalvo00}. The thermal Comptonized component
contributes more than 80 per cent of the total flux.  The spectral
evolution of this source has been previously studied using {\it BeppoSAX}
data \citep{disalvo00}. However, {\it BeppoSAX} data revealed only
the HB and NB of the source. During our observations, source has shown
extended FB making it possible to study the spectral behaviour of the
source along a complete Z-track.

As the source GX 17+2 moves along the HB and then down the NB, the
optical depth $\tau$ of the corona decreases which is in accordance
with the previously observed behaviour exhibited by the source GX
17+2 \citep{disalvo00}. The increase in the optical depth along the
FB is similar to that observed in the Z-source LMC X-2 previously
\citep{agrawal09}. The decrease in optical depth along the NB and then
increase along the FB has also been previously reported in the Z-source
GX 349+2 \citep{agrawal03}. The electron temperature $kT_e$ in the HB
and NB is $\sim$ 3.2 keV. Similar values of $kT_e$ was found during {\it
BeppoSAX} observation of the source in the HB and NB \citep{disalvo00}.
In the FB, the $kT_e$ is slightly lower.

The evolution of the Comptonized spectrum can be explained in terms
of increasing accretion rate scenario. In this scenario, as the soft
photon supply from neutron star surface/disc increases, the part of the
material from the corona interacts with these photons and settles down
in the disc. This causes the reduction in the optical depth of the
corona. This explains the observed behaviour of the source along the HB
to NB. Similarly, as the accretion rate further increases radiation
pressure pushes a part of the material from the accretion disc into the
corona causing it to become optically thick (see also \citealt{agrawal03}).
Comptonization parameter $y$, which is a measure of relative energy
gain, decreases along the HB and then upto lower part of the NB (from
NB1 to NB3) causing spectrum to become softer. Similarly increase in
the $y$-parameter causes the source to move up on the FB. The reported
variation in the $y$ parameter is in accordance with that observed in
the source GX 349+2 \citep{agrawal03} and LMC X-2 \citep{agrawal09}.

{\bf Model 3} which includes a reflection component from the ionized disc
provides equally good fit to the spectra of the source. The reflection
component does not require a relativistic smearing, suggesting that iron
line is narrow. This is supported by the fact that inner rim of the disk
is truncated away from the compact object ($10-45 R_g$). The reflection
fraction varies between $0.3-0.5$ and {\bf attains the highest values}
near the NB-FB vertex (see Table 4). We also note that the equivalent
width of the iron line derived using {\bf Model 1} is correlated with the
reflection fraction of {\bf Model 3} which further supports that iron line
is due to reflection of Comptonized emission from the truncated disc.
{\bf Model 4}, which includes the reflecction of the boundary layer
emission from the ionized accretion disk, {\bf is} also a good description
of the spectra. Here, we have also used a reletivistic convolution which
suggests that the accretion disc is very far from the compact object
($10 - 20$ $R_{ISCO})$. This suggests that the space between the compact
object and inner rim of accretion disc should be filled with some hot
plasma. Hence, {\bf Model 3} which invokes reflection of the Comptonized
emission emitted by this plasma by a ionized accretion disc seems to be
a more promising scenario. Also reflection fraction (from {\bf  Model 3}) is much better correlated (correlation coeffficient = 0.94)  with equivalent width of iron line compared to reflection fraction obtained using BBREFL model (correlation coefficient = 0.63), again suggesting that {\bf Model 3} is a better description.

Modelling of the relativistic iron line with {\it diskline} and reflection
models suggests that the accretion disc is truncated away from the neutron
star. However, modelling of the reflection signature observed with the
{\it NuSTAR} data suggests that the  accretion disc in this source lies
close to the inner most stable circular orbit (ISCO) \citep{ludham17},
although these authors do not mention the spectral branch where they have
detected the reflection feature.
  
A normal branch oscillation (NBO) with frequency 7.42$\pm$0.29 Hz is
detected at the NB2 section of the NB. If we consider {\bf Model 1} as
correct description of the spectra, the thermal Comptonized emission
is the only dominant component in the $3.0-80.0$ keV energy band. It is
more probable that an optically thick central corona is producing this
emission component by up-scattering of the soft seed photons supplied by
the neutron star surface. Hence, it seems that the radiation
feedback mechanism with this hot central corona is probably producing
the NBO (see \citealt{Fortner89}) observed in this source.

\section*{Acknowledgements}
We thank anonymous reviewer for providing useful suggestions which improved the quality of the paper.
This research has made use of the data obtained through GT phase of {\it
AstroSat} observation. Authors thank GD, SAG; DD, PDMSA and Director, URSC
for encouragement and continuous support to carry out this research. This
work has used the data from the LAXPC Instruments developed at TIFR,
Mumbai and the LAXPC POC at TIFR is thanked for verifying and releasing
the data via the ISSDC data archive. We thank the AstroSat Science Support
Cell hosted by IUCAA and TIFR for providing the LaxpcSoft software which
we used for LAXPC data analysis. 

\end{document}